\documentclass[aps,prl,twocolumn,reprint,floatfix, superscriptaddress]{revtex4-2}

\usepackage{bm}

\usepackage{graphicx}
\graphicspath{{figures/}}

\usepackage[T1]{fontenc}
\usepackage{amsmath, amssymb}
\usepackage{physics}
\usepackage{dsfont}
\usepackage{soul}

\usepackage[colorlinks=true,linkcolor=blue,urlcolor=blue,citecolor=blue]{hyperref}

\usepackage{xcolor}

\begin{document}
	
	
	\newcommand{\mytitle}{Quantum control of hole spin qubits in double quantum dots}
	
	\title{\mytitle}
	
	\author{D.~Fern\'andez-Fern\'andez}
	\affiliation{Instituto de Ciencia de Materiales de Madrid ICMM-CSIC, 28049 Madrid, Spain}
	
	\author{Yue Ban}
	\affiliation{Department of Physical Chemistry, University of the Basque Country UPV/EHU, Bilbao, Spain}
	\affiliation{EHU Quantum Center, University of the Basque Country UPV/EHU, Leioa, Biscay, Spain}
	\affiliation{TECNALIA, Basque Research and Technology Alliance (BRTA), 48160 Derio, Bizkaia, Spain}
	
	\author{G.~Platero}
	\email{gplatero@icmm.csic.es}
	\affiliation{Instituto de Ciencia de Materiales de Madrid ICMM-CSIC, 28049 Madrid, Spain}

	
	\begin{abstract}
		Hole spin qubits in semiconductor quantum dots (QDs) are promising candidates for quantum information processing due to their weak hyperfine coupling to nuclear spins, and to the strong spin-orbit coupling which allows for rapid operation time. We propose a coherent control on two heavy-hole spin qubits in a double QD by a fast adiabatic driving protocol, which helps to achieve higher fidelities than other experimentally commonly used protocols as linear ramping, $\pi$-pulses or Landau-Zener passages. Using fast quasiadiabatic driving via spin-orbit coupling, it is possible to reduce charge noise significantly for qubit manipulation and achieve high robustness for the qubit initialization. We also implement one and two-qubit gates, in particular, NOT, CNOT, and SWAP-like gates, of hole spins in a double QD achieving fidelities above $99\%$, exhibiting the capability of hole spins to implement universal gates for quantum computing.
	\end{abstract}
	
	
	\maketitle
	
	\textit{Introduction.} Fast and precise control of a large number of qubits is required for the implementation of quantum algorithms and hardware to realize quantum computing. As one of the pillars in the development of quantum technologies \cite{Tarucha1996,Loss1998,Ciorga2000,Hanson2007,Kloeffel2013}, spin qubits in quantum dots (QDs) \cite{Busl2013,Sanchez2014,Pico2019} present long coherence time \cite{Veldhorst2015} and compatibility in semiconductor manufacturing technology \cite{Zwerver2021}. Recent progress with a fidelity exceeding $99\%$ overcomes the barrier for two-qubit gate control \cite{Mills2021, Madzik2022, Noiri2022, Xue2022}, which signifies that semiconductor qubits possess promising potential applications in the era of noisy intermediate-scale quantum devices. Great effort is currently being devoted to the investigation of hole spin qubits in QDs \cite{Bulaev2005,Heiss2007,Pribiag2013,Prechtel2016,Bogan_2017,Bogan2018,Liles2018,Bogan_2019,Hendrickx2020,Hendrickx_2020,Mutter_2021,Froning_2021,van_Riggelen_2021,Hendrickx_2021,Wang2022,Wang2021}, owing to their long coherence time resulting from the weak hyperfine coupling to nuclear spins \cite{Prechtel2016}, and rapid operation time \cite{Kloeffel2011,Szumniak2012,Bogan2018} due to the inherently strong spin-orbit coupling (SOC). In particular, Ge QDs \cite{Watzinger2018,Vukusic2018,Hendrickx2018,Froning_2021b} have high hole mobility and a strong Rashba SOC \cite{Kloeffel2011,Crippa2018}, which facilitates electrical drive for fast qubit operations. In GaAs QDs however, due to the lack of bulk inversion symmetry, Dresselhaus SOC also plays an important role. All of these allow for a wide range of tunable properties, leading to highly scalable, easily addressable, and fast hole spin qubits. However, all-electric control related to strong SOC could induce a higher susceptibility to charge noise, which is detrimental to the robustness of hole spin qubits.
	
	Landau-Zener (LZ) protocols have been developed to manipulate charge and spin electron qubits \cite{Ribeiro_2009,Ribeiro_2010,Petta_2010,Nichol2015,Gomez_2019}. Low ramping velocity and adiabatic pulses allow for the transition along the instantaneous eigenstate \cite{Greentree2004,Huneke2013}, but are prone to decoherence, whereas fast ramping causes transitions between instantaneous eigenstates. Pulses in different shapes like ``double hat'', ``convolved'', and ``trapezoid'' \cite{Ribeiro2013} solve this trade-off to some extent. However, high accuracy needs further tuning of the parameters whose modulation can allow for the best compromise. To this end, shortcuts to adiabaticity (STA), a framework that allows reducing the duration of slow adiabatic processes \cite{Chen2010,Ruschhaupt2012,Huneke2013,Torrontegui2013,Li2018}, is believed to solve the above issue. It has been applied for fast and robust control of electron spin qubits in a single \cite{Ban_2012} and a double QD (DQD) \cite{Khomitsky_2012,Ban_2014}, and for electron transfer in a long QD array \cite{Ban_2018, Ban_2019}. Moreover, dynamical sweet spots have been investigated to increase the spin qubit decoherence time \cite{PicoCortes2021}. Also, non-adiabatic geometric phases, such as the Aharonov-Anandan phase \cite{Aharonov1987}, have been used to construct electron spin gates in semiconductor QDs \cite{Zhang2020,Chen2022} with reduced evolution time. 
	
	\begin{figure}[!htbp]
		\centering
		\includegraphics[width=\linewidth]{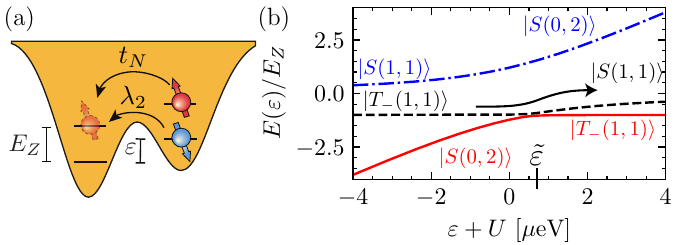}
		\caption{(a) Scheme of a DQD populated with two holes with level detuning $\varepsilon$. $t_N$ is the spin-conserving tunneling rate between dots, while the spin-flip tunneling rate $\lambda_2$ is a consequence of the SOC. The magnetic field perpendicular to the dots plane produces a Zeeman splitting $E_Z$. (b) Energy-level diagram of the system. Basis states at large and low detunings are indicated. The $S-T_-$ anticrossing is located at $\tilde{\varepsilon}$. $B=15$ mT, $U=2$ meV, $t_N=1\; \mu$eV, $\lambda_2=0.1\; \mu$eV.}
		\label{fig:system}
	\end{figure}
	
	In this work, we propose a high-fidelity control protocol for the hole spin singlet-triplet qubit where both Rashba and Dresselhaus SOC play a prominent role. By fast quasi-adiabatic (FAQUAD) approach \cite{MartinezGaraot2015}, the detuning between energy levels $\varepsilon(t)$ is used as the control parameter so that the state evolution is along the adiabatic state as fast as possible, avoiding diabatic transitions. We compared the results obtained by FAQUAD and other protocols, such as a linear ramp, the LZ protocol, a $\pi$-pulse, and a reduced version of FAQUAD known as Local Adiabatic (LA). Each of them may exhibit its advantages depending on the working scenario. However, we show that the proposed FAQUAD protocol for quantum control of a hole spin qubit allows achieving higher fidelities. Initializing the qubit at an arbitrary state, hole spin one-qubit and two-qubit gates such as CNOT and SWAP gates, achieve fidelities beyond the fault-tolerance error correction threshold of 0.99 \cite{Raussendorf_2007,Wang_2011,Fowler_2012}. To the best of our knowledge, there are no other proposals in the literature for the design of two-qubit gates for hole spin singlet-triplet qubits. In this work, we focus on GaAs QDs where both Rashba and Dresselhaus interactions are present. However, the theoretical framework is general and therefore applicable to other materials such as Si or Ge where only Rashba SOC is present.\\
	
	\textit{Model.} We consider a planar DQD populated with two heavy holes (HH), Fig.~\ref{fig:system} (a). The total Hamiltonian can be written as $H_0 = H_\text{DQD} + H_\text{B} + H_\text{SOC}$, where the first term reads
	\begin{equation}
		\begin{split}
			H_\text{DQD} =& \sum_{i=L,R}\varepsilon_i c_i^\dagger c_i + U\sum _{i=L,R}n_{i\uparrow}n_{i\downarrow} \\
			& - t_N\sum_{\sigma=\uparrow, \downarrow}(c_{L\sigma}^\dagger c_{R\sigma}+\text{H.c.}).
		\end{split}
	\end{equation}
	Here $\varepsilon_i$ with $i=L, R$ is the energy level of the left and right dots, respectively, $U$ is the intradot Coulomb energy, and $t_N$ the spin-conserving tunneling rate between adjacent dots. The operator $c_{i\sigma}$ ($c_{i\sigma}^\dagger$) is the annihilation (creation) of a hole state with spin $\sigma$ in the dot $i$. The applied magnetic field with intensity $B$, perpendicular to the dots plane, results in a Zeeman splitting $H_\text{B} = \frac{1}{2}g^*\mu_B B \sum_{i=L,R}(n_{i\uparrow} - n_{i\downarrow})$, where $g^*$ is the effective Landé factor ($g^*=1.45$ for holes in GaAs) and $\mu_B$ the Bohr magneton. Here we assume a homogeneous magnetic field and g-factor, while the effect of an inhomogeneous g-factor is analyzed in the Supplemental Material \cite{SuppMat}. Finally, SOC is given by the term
	\begin{equation}
		\begin{split}
			H_\text{SO}= & \;i\alpha E_\perp (\sigma_+p_-^3 - \sigma_-p_+^3) \\
			&- \beta(\sigma_+p_-p_+p_- + \sigma_-p_+p_-p_+).
		\end{split}
		\label{eq:H_SOC}
	\end{equation}
	The ladder operators are defined as $\sigma_\pm = (\sigma_x\pm i\sigma_y)/\sqrt{2}$, the momentum operator $p_\pm =p_x\pm ip_y$ given by $\vec{p}=-i\hbar \vec{\nabla} + e^*/c\vec{A}$, with $e^*$ the hole effective electric charge, $c$ the speed of light, and $\vec{A}$ the magnetic vector potential. Eq.~(\ref{eq:H_SOC}) represents the Rashba SOC ($\alpha$) due to the structure inversion asymmetry, controlled by the effective electric field $E_\perp$ produced by the accumulation gate, and Dresselhaus SOC ($\beta$) due to the bulk inversion asymmetry. SOC in two-dimensional hole gases was analyzed in III-V and Si-based heterojunctions in Ref.~\cite{Marcellina2017}, where it was shown that higher-order terms in the wave vector contained in the heavy hole spin splitting are sizable. In GaAs QDs, cubic Rashba and Dresselhaus SOC are the dominant terms, while in systems like Ge and Si nanowires and elongated QDs, the spin-orbit is modeled by the direct Rashba SOC \cite{Bosco_2021,Adelsberger_2022}. Furthermore, depending on the QD configuration, linear contributions to the SOC could be important \cite{Luo2010,High2013,Durnev2014,Liu2022}. Nevertheless, in all these cases SOC can be included in a phenomenological model as a spin-flip term.
	
	Since we consider a closed system, it is appropriate to work within the molecular basis of singlets: $\ket{S(1,1)}\equiv(\ket{\uparrow,\downarrow}-\ket{\downarrow,\uparrow})/\sqrt{2}$, $\ket{S(0, 2)}\equiv\ket{0, \uparrow\downarrow}$ and $\ket{S(2, 0)}\equiv\ket{\uparrow\downarrow, 0}$, and triplets $\ket{T_0(1, 1)}\equiv(\ket{\uparrow,\downarrow}+\ket{\downarrow,\uparrow})/\sqrt{2}$, $\ket{T_-(1, 1)}\equiv\ket{\downarrow,\downarrow}$ and $\ket{T_+(1, 1)}\equiv \ket{\uparrow, \uparrow}$. We write the spin-flip tunneling matrix element between the polarized triplet states and the double-occupied singlet state as $\bra{T_\pm(1, 1)}H_\text{SOC}\ket{S(0, 2)} = \lambda_2$, and $\bra{T_\pm(1, 1)}H_\text{SOC}\ket{S(1, 1)} = 2\lambda_2S / \sqrt{2} = \lambda_1$ for the single-occupied singlet state. Here we define the overlap between the wave functions in each dot as $S\equiv \braket{L}{R}=\exp(-d^2/2l^2)$, where $d$ is the distance between dots, and $l$ the extent of the wave function centered at each dot. The matrix element $\lambda_2$ also depends on these two parameters. The explicit expression showing this dependence is given in the supplemental material \cite{SuppMat}. With the experimental parameters for GaAs QDs proposed in Ref.~\cite{Bogan2018}, we obtain the relation $\lambda_1/\lambda_2 \sim 1/100$.
	
	Under a constant magnetic field $B$ the unpolarized triplet state $\ket{T_0(1, 1)}$ does not interact with any other state. The anticrossing between $\ket{T_+(1,1)}$ and the singlet states is located in a detuning $\varepsilon \equiv \varepsilon_R - \varepsilon_L$ close to $\varepsilon + U \sim -2t_N^2/E_Z+E_Z$. As we will see below, our working regime is far from this detuning so $\ket{T_+(1,1)}$ does not interact with other states in our case, and we can neglect its contribution. Furthermore, if the detuning is large enough ($\varepsilon>-U$) the double-occupied singlet state $\ket{S(2, 0)}$ is far apart in energy and does not play any role. Then, we write the matrix form of the total Hamiltonian $H_0$ in the basis $(\ket{T_-(1,1)}, \ket{S(0,2)}, \ket{S(1,1)})$ as
	
	\begin{equation}
		H_0 =\mqty(-E_Z & \lambda_2 & \lambda_1 \cr
		\lambda_2^* & \varepsilon +U & -\sqrt{2}t_N \cr
		\lambda_1^* & -\sqrt{2}t_N & 0 \cr)\; ,
		\label{eq:Hamiltonian_DQD}
	\end{equation}
	with the Zeeman splitting in each dot $E_Z$, and $\varepsilon_L + \varepsilon_R = 0$. The instantaneous eigenenergies are shown in Fig.~\ref{fig:system} (b). Due to the finite spin-conserving tunneling rate $t_N$, the singlets form hybridized states
	\begin{equation}
		\begin{split}
			\ket{S_G} &= \cos\Omega/2 \ket{S(1, 1)} + \sin\Omega/2\ket{S(0, 2)}, \\
			\ket{S_E} &= -\sin\Omega/2 \ket{S(1, 1)} + \cos\Omega/2\ket{S(0, 2)}, \\
		\end{split}
	\end{equation}
	where $\tan\Omega = 2\sqrt{2}t_N / (\varepsilon + U)$. We encode our computation basis with the triplet state and the ground hybridized singlet state. However, the transition between the charge configurations $(1, 1)$ and $(0, 2)$, leads to charge fluctuation. This is why we will try to reduce the population of $\ket{S(0, 2)}$ as much as possible. The anticrossing between singlet states is located at $\varepsilon + U$ = 0. Far from this point, for $\varepsilon + U \gg t_N$, the ground hybridized singlet state is $\ket{S_G} \sim \ket{S(1, 1)}$. It is this regime in which we are interested since charge noise is significantly reduced. The anticrossing between the triplet and the ground hybridized singlet state is located at $\tilde{\varepsilon} + U \sim 2t_N^2 / E_Z - E_Z$. Then, we need that this anticrossing fulfills $\tilde{\varepsilon} + U \gg t_N$, which is verified in the limit $t_N \gg E_Z$, being this configuration the ideal one to avoid a significant contribution of $\ket{S(0, 2)}$. The typical value for the spin-conserving tunneling rate is close to $t_N\sim 5 \;\mu$eV, while a reasonable magnetic field of $B\sim5$~mT corresponds to a Zeeman spiting $E_Z\sim0.4\;\mu$eV, verifying the above condition. In particular, if $t_N > E_Z/\sqrt{2}$ the anticrossings between the singlet state and the triplet states $\ket{T_-(1,1)}$ and $\ket{T_+(1,1)}$ are far apart from each other, so the approximation of neglecting $\ket{T_+(1, 1)}$ is justified. If other excited states are close to the working point, more elaborate protocols should be used to maintain high fidelity in the operations to avoid a possible leakage out of the computation basis. In particular, the speed of the protocol will be reduced in order to avoid transitions to these states.\\
	
	\textit{Transfer $S(1,1)-T_{-}(1,1)$.} We investigate the design of a detuning pulse in order to pass from $\ket{T_-(1,1)}$ to $\ket{S(1,1)}$ along the instantaneous state $\ket{\psi_1}$, where $H_0\ket{\psi_1(\varepsilon)}=E_1(\varepsilon)\ket{\psi_1(\varepsilon)}$, whose instant eigenenergy $E_1(\varepsilon)$ is represented by the dashed black line in Fig.~\ref{fig:system} (b). There are different approaches to design the detuning pulse between the dots, the only driving parameter in our case. A liner ramp of the detuning between the initial and final time gives the adiabatic evolution when the time derivative of the detuning, i.e., the driving speed, is small enough $(\varepsilon(t_f) - \varepsilon(0)) / t_f \ll \lambda_2$ so the adiabatic condition is fulfilled during the protocol. To reduce the total time, one can consider an LZ-type protocol composed of a linear ramp with total time $t_\text{raise}$, a waiting time $t_w$, and another linear ramp returning to the original point. The total time of the protocol is given by $t_f = 2t_\text{raise} + t_w$. During the waiting time, the state acquires a phase interfering destructively or constructively during the returning linear ramp, resulting in a complete transfer. Another proposal for quantum control is the $\pi$-pulse. In this protocol, all the parameters of the system are constant over time, with the detuning fixed at the anticrossing $\tilde{\varepsilon}$. By changing the total time of the protocol, the system undergoes Rabi oscillations between both states. If the waiting time is a multiple of $\pi/\hbar f(\lambda_2)$, where $f(\lambda_2)$ is the SOC energy gap, a transfer between states is achieved. However, this protocol only works properly when the system can be effectively reduced to a two-level system. If not, multiple $\pi$-pulses must be employed in each of the anticrossings \cite{Holthaus1994}. In this work, we propose to use a protocol, not yet experimentally implemented in semiconductor quantum dots, based on shortcuts to adiabaticity, in order to improve the fidelity of the established protocols mentioned above.
	
\begin{figure}[!htbp]
		\centering
		\includegraphics[width=\linewidth]{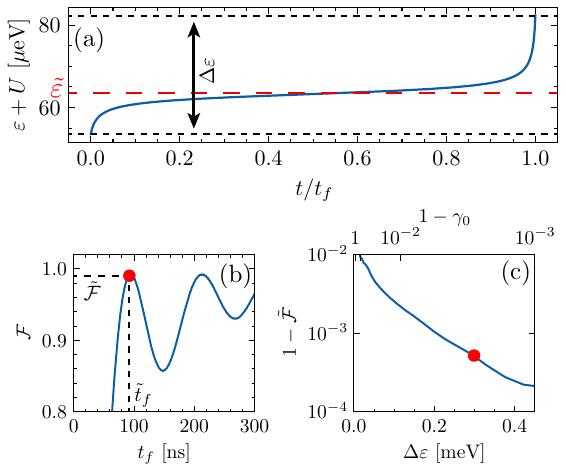}
		\caption{(a) Detuning pulse shape obtained with FAQUAD. The boundary conditions for the detuning define the detuning range as $\Delta\varepsilon =\varepsilon(t_f) - \varepsilon(0)$. The dashed red line marks the location of the anticrossing. (b) Fidelity of the $\ket{T_-(1, 1)}\rightarrow \ket{S(1, 1)}$ transfer versus the total time of the protocol. The red dot marks the first peak in the fidelity $\tilde{\mathcal{F}}$ which is obtained for a total time of $\tilde{t}_f$. The detuning range used is $\Delta \varepsilon = 0.3$ meV. (c) Infidelity of FAQUAD versus the detuning range $\Delta\varepsilon$. The red dot corresponds to the fidelity shown in (b). $B=10$ mT, $U=2$ meV, $t_N=5\; \mu$eV, $\lambda_2=0.1\; \mu$eV, $\lambda_1 = \lambda_2 / 100$.}
		\label{fig:range_epsilon}
	\end{figure}

	We investigate the FAQUAD \cite{MartinezGaraot2015} protocol, which efficiently allows to speed up the transfer far from the anticrossings where diabatic transitions to excited states have a very low probability, and reduce the speed otherwise. The adiabaticity condition for a $N$-level system can be written as 
	\begin{equation}
		c = \hbar \sum_{k\neq i}^N\abs{\frac{\bra{\phi_i(t)}\partial_t H(t)\ket{\phi_k(t)}}{\left[E_i(t)-E_k(t)\right]^2}},
		\label{eq:adiabatic_condition}
	\end{equation}
	where $\ket{\phi_k(t)}$ are the instantaneous eigenstates and $E_k(t)$ the corresponding eigenenergies. The system is initialized in the eigenstate given by $\ket{\Psi(0)} = \ket{\phi_i(0)}$. If the dynamics is slow enough, that is, $c\ll 1$, at the end of the protocol the system will remain in the same eigenstate $\ket{\Psi(t_f)} = \ket{\phi_i(t_f)}$. We assume that the system is driven by a single parameter, named $\varepsilon$, so we can write $\partial_tH = \dot{\varepsilon}\partial_\varepsilon H $, with $\dot{\varepsilon}\equiv\partial_t\varepsilon$ the driving speed. Imposing boundary conditions for the driving parameters (see below) we can rewrite the adiabatic condition in Eq.~(\ref{eq:adiabatic_condition}) as 
	\begin{equation}
		c = \frac{\hbar}{t_f}\sum_{k\neq i}^N\int_{\varepsilon(0)}^{\varepsilon(t_f)}d\varepsilon\abs{\frac{\bra{\phi_i(\varepsilon)}\partial_\varepsilon H(t)\ket{\phi_k(\varepsilon)}}{\left[E_i(\varepsilon)-E_k(\varepsilon)\right]^2}}.
	\end{equation}
	The above integral has no analytical solution for a general system. However, it can be easily solved by numerical methods. Once the value of the adiabatic constant $c$ is known, we can solve the following differential equation to obtain the time-dependent driving parameter
	\begin{equation}
		\dot{\varepsilon} = \frac{c}{\hbar}\sum_{k\neq i}^N\abs{\frac{\left[E_i(\varepsilon)-E_k(\varepsilon)\right]^2}{\bra{\phi_i(\varepsilon)}\partial_\varepsilon H(t)\ket{\phi_k(\varepsilon)}}}.
		\label{eq:driving_parameter}
	\end{equation}
	The solution of the above equation gives as a result a driving parameter that ensures a constant value of $c$ during the transfer. The dynamic evolves fast if the control parameter is far from the avoided anticrossings, where no diabatic transitions are possible. Near the anticrossings, the driving slows down to ensure the adiabatic condition, staying in the instantaneous eigenstate, Fig.~\ref{fig:range_epsilon} (a). we define the fidelity of the protocol as the total population of the single occupied singlet state at the final time $\mathcal{F}\equiv |\braket{S(1,1)}{\Psi(t_f)}|^2$. In Fig.~\ref{fig:range_epsilon} (b) we show how the fidelity depends on the total time of the protocol. As $t_f$ increases the dynamic is closer to the adiabatic regime, i.e, $c\ll 1$. The characteristic of the FAQUAD protocol is the ondulatory behavior of fidelity, tending asymptotically to the value of unity. We define the first peak in the fidelity as $\tilde{\mathcal{F}}$, which is reached with a total time of $\tilde{t}_f$.
	
	The boundary conditions for the driving parameter have a significant effect on the final result of the transfer. We start the dynamics initializing the system in the triplet state $\ket{\Psi(t=0)}=\ket{T_-(1,1)}$. However, this state does not exactly corresponds to an instant eigenstate $\ket{\phi_i}$ with $\abs{\braket{T_-(1, 1)}{\phi_i(\varepsilon(0))}}^2 = 1 - \gamma_0$ for $0<\gamma_0<1$. Ideally, the protocol would start with a detuning such that $\gamma_0 = 0$, but this is only possible at the limit $\varepsilon(0) \rightarrow -\infty$. Then, working with a moderate value of $\varepsilon(0)$, $\gamma_0$ acquires a finite value. The same discussion is valid for the final detuning at which $\abs{\braket{S(1, 1)}{\phi_i(\varepsilon(t_f))}}^2 = 1 - \gamma_f$, but now the value of $\gamma_f=0$ is only reached in the limit of large detuning $\varepsilon(t_f) \rightarrow \infty$. We study the dependence of $\tilde{\mathcal{F}}$ on the value of the boundary conditions fixing $\gamma_0=\gamma_f$. In Fig.~\ref{fig:range_epsilon} (c) we plot the infidelity at the first peak versus the detuning $\Delta\varepsilon\equiv \varepsilon(t_f) - \varepsilon(0)$. As we increase the range of detuning, the values of $\gamma_0$ and $\gamma_f$ decrease, resulting in higher transfer fidelity. We find that the fidelity increases exponentially with the detuning range. Using a large detuning range is a promising way to achieve ultra-high transfer fidelities. Furthermore, the total time needed to reach the first peak remains nearly constant when increasing the detuning range (not shown here). During the rest of the article we fix the boundary conditions such that $\gamma_0=\gamma_f=0.01$, obtaining a moderate value for the detuning $\Delta\varepsilon \sim 30\; \mu$eV.
	
\begin{figure}[!htbp]
		\centering
		\includegraphics[width=\linewidth]{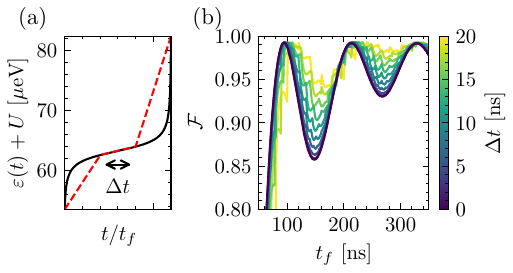}
		\caption{(a) Ideal pulse shape (solid black line), and a discretized pulse with certain time resolution $\Delta t$. (b) Fidelity for the $\ket{T_-(1, 1)}\rightarrow \ket{S(1, 1)}$ transfer with FAQUAD against the total time of the protocol for different values of the time resolution $\Delta t$. The limit $\Delta t \rightarrow 0$ (darker colors) denotes the limits of the ideal pulse. $g = 1.35$, $B = 10$~mT, $t_N = 5\; \mu$eV, $\lambda_2 = 0.1\; \mu$eV, and $U = 2$~meV.}
		\label{fig:finite_dt_all}
	\end{figure}

	Besides the detuning range, an important thing to keep in mind for a possible experimental implementation is the pulse shape. With FAQUAD we obtain a control parameter pulse shape with sharp edges at both the beginning and the end of the protocol. In order to mimic a pulse that could be experimentally implemented, we divide the ideal pulse shape for the detuning into a series of linear ramps with an individual duration of $\Delta t$, see Fig.~\ref{fig:finite_dt_all} (a). 
	The pulse is then divided into a total of $t_f / \Delta t$ linear ramps, recovering the ideal pulse at $\Delta t = 0$. In Fig.~\ref{fig:finite_dt_all} (b) we plot the fidelity for the $\ket{T_-(1,1)}\rightarrow \ket{S(1,1)}$ transfer versus the total time of the protocol using the ideal pulse, along with more realistic pulses with different time resolution $\Delta t$. In all cases, the maximum fidelity is $\mathcal{F} > 0.99$, even for values as high as $\Delta t = 20$ ns. We demonstrate that the pulses obtained with FAQUAD are robust against a finite time resolution of the control parameter, being a potential candidate for the manipulation of spin qubits in semiconductor QDs.
	
\begin{figure}[!tp]
		\centering
		\includegraphics[width=\linewidth]{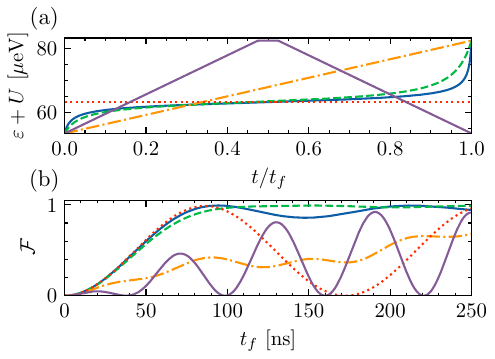}
		\caption{(a) Pulses shapes for the state transfer $T_-(1,1)\rightarrow S(1,1)$, designed by FAQUAD (solid blue), LA (dashed green), linear (dotted-dashed orange), $\pi$-pulse (dotted red), and the LZ-type protocol consisting in two linear ramps and a waiting time (solid purple). (b) Fidelity for the different protocols against the total time. $B = 10$ mT, $U=2$ meV, $t_N=5\; \mu$eV, $\lambda_2=0.1\; \mu$eV, $\lambda_1 = \lambda_2 / 100$, and $t_w = 0.05t_{\text{raise}}$.}
		\label{fig:pulses_alternatives}
	\end{figure}

	A simplified version of FAQUAD known as Local Adiabatic (LA) \cite{Roland_2002,Richerme_2013, SuppMat} also gives rise to outstanding results. This protocol requires less information about the system if in Eq.~(\ref{eq:adiabatic_condition}) the next condition $\bra{\phi_i(t)}\partial_t H(t)\ket{\phi_k(t)} = 1$, $\forall k\neq i$, is considered. This assumption, even if not verified at all times, highly simplifies the protocol, which no longer needs information about the eigenstates. It could be beneficial for its implementation where a precise characterization of all the parameters can be challenging (see the Supp. Material for more details). Furthermore, the fidelity obtained with this protocol is comparable to the results obtained with FAQUAD. Fig. \ref{fig:pulses_alternatives} (a) shows the pulse derived from each of the mentioned protocols. Among all protocols, the highest fidelity is obtained by FAQUAD and LA, with the maximum value $\mathcal{F}_\mathrm{max} = 0.993$ (Fig.~\ref{fig:pulses_alternatives} (b)). FAQUAD reached this value at a smaller $t_f$ than LA, while the latter is less sensitive to a deviation in the final time $t_f$. Using a $\pi$-pulse, we observe typical Rabi oscillations, but with $\mathcal{F}<0.99$ at all final times. To understand the low fidelity obtained by performing a linear protocol, we can assume that we have a two-level system and that we can apply the Landau-Zener formula $\mathcal{F}^\text{LZ}=1-\exp(-2\pi \lambda_2^2/\hbar\nu)$, where $\nu=\Delta\varepsilon / t_f$ is the speed of the linear ramp. We find that the detuning range is proportional to the spin-conserving tunneling rate $\Delta\varepsilon\propto t_N$. Using experimental parameters for HH in GaAs the SOC $\lambda_2/t_N=0.02$ \cite{Jirovec2022} is small. Using this parameters in the Landau-Zener formula we obtain low fidelity even for large times $\mathcal{F}^\text{LZ}(t_f=250\;\text{ns})\sim 0.6$. This value is close to that obtained by numerical methods for the linear ramp (Fig.~\ref{fig:pulses_alternatives} (b), orange line). Applying a protocol consisting of two linear pulses plus a waiting time in between substantially improves the results of a linear ramp. However, the maximum fidelity obtained with this pulse is still below the proposed FAQUAD protocol, which proves to be a better alternative than those protocols usually considered in experiments. Since the best results for state transfer are obtained with FAQUAD, we focus on this protocol and compare its feasibility with the other established ones.
	
\begin{figure}[!tpb]
		\centering
		\includegraphics[width=\linewidth]{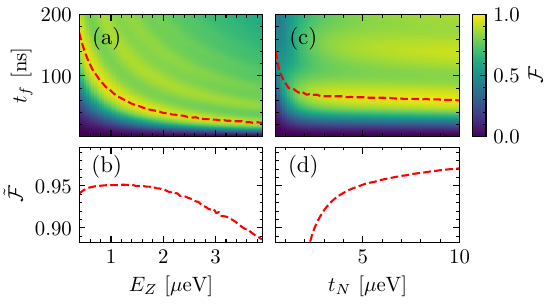}
		\caption{(a),(c) Fidelity, considering the FAQUAD protocol, for the transfer $\ket{T_-(1, 1)}\rightarrow \ket{S(1, 1)}$ in presence of charge noise and spin dephasing as a function of (a) $E_Z$ and $t_f$ with $t_N = 5\; \mu$eV; (c) $t_N$ and $t_f$ with $E_Z=1.17\; \mu$eV. (b),(d) $\tilde{\mathcal{F}}$ corresponding to the dashed red lines shown in (a, c). $U=2$ meV, $\lambda_2=0.02 t_N$, $\lambda_1 = \lambda_2 / 100$, $\gamma_2=10^{-2}$ ns$^{-1}$, $\Gamma_\mathrm{sd}=10^{-4}$ ns$^{-1}$.}
		\label{fig:charge_noise}
	\end{figure}

	In order to study the effect of the charge noise on the state transfer, we solve the master equation $\dot{\rho} = -i/\hbar [H_0,\rho] + \sum_i\left(L_i\rho L_i^\dagger - 1/2 \left\{L_i^\dagger L_i, \rho\right\}\right)$, where the Lindblad operator is given by $L_i = (\sqrt{\Gamma_\mathrm{ch}} + \sqrt{\Gamma_\mathrm{sd}}) \sigma_i$, and $\sigma_i$ are the two diagonal Gell-Mann matrices. Pure dephasing is mainly caused by charge noise when $\ket{S(0, 2)}$ is populated during the transfer. It leads to a dephasing strength $\Gamma_\mathrm{ch}(\varepsilon) = \gamma_2\abs{\braket{S(0, 2)}{\psi_1(\varepsilon)}}^2$ \cite{Ribeiro2013}. Furthermore, we also include an extra spin dephasing term ($\Gamma_\mathrm{sd}$) due to the spin-orbit mixing of the HH states interacting with phonons \cite{Bulaev2005}, and to the hyperfine interaction. This spin dephasing term is assumed to be constant at all detuning. Fig.~\ref{fig:charge_noise} (a-b) shows the fidelity of FAQUAD in terms of $t_f$ and $E_Z$ in presence of charge noise and spin dephasing. In the limit of a large magnetic field, the total time needed to reach the first peak in the fidelity is much smaller than the spin dephasing strength $\tilde{t}_f \ll 1 / \Gamma_\mathrm{sd}$. However, the crossing between the singlet and the triplet states is located at $\tilde{\varepsilon} + U < 0$. In this region the ground state $\ket{S_G} \sim \ket{S(0, 2)}$ and charge noise during the dynamics is the dominant noise source. The large population of $\ket{S(0, 2)}$ results in low transfer fidelities. In the other limit, working with low magnetic fields, the relevance of $\ket{S(0, 2)}$ decreases, while the total time $\tilde{t}_f$ that is needed to obtain a complete state transfer increases. Even if charge noise is highly suppressed in this limit, $\tilde{t}_f$ is high enough such that spin dephasing is relevant, resulting in a decrease of the fidelity $\tilde{\mathcal{F}}$. Therefore, there is a compromise between these two limits for the magnetic field intensity. With the parameters considered, the maximum fidelity is obtained at $B\sim 15$ mT, i.e., $E_Z\sim 1.17\; \mu$eV. This value corresponds to the optimal point of operation for the state transfer.
	
	We also analyze the dependence of $\mathcal{F}$ on the spin-conserving tunneling rate, Fig.~\ref{fig:charge_noise} (c-d) in the presence of charge noise and spin dephasing. Here we fix the ratio between the spin-flip and spin-conserving tunneling rates at $\lambda_2/t_N = 0.02$. As the tunneling rate increases, the crossing point moves farther from $\tilde{\varepsilon} + U = 0$, so that the ground hybrid singlet state is given by $\ket{S_G} \sim \ket{S(1, 1)}$. Then, $\ket{S(0, 2)}$ is not populated during the transfer resulting in low charge noise. In this case, increasing $t_N$ has low effect on the total time $\tilde{t}_f$. However, working with large $t_N$ results in a increase of the detuning range $\Delta \varepsilon\equiv \varepsilon(t_f) - \varepsilon(0)$ needed to obtain the same boundary condition of $\gamma_0$ and $\gamma_f$. If the initial of the final detuning is high enough, surpassing the lead barriers, new particles can enter the system, modifying the results shown here. Therefore, for a practical scenario in an experimental device, a moderate $t_N$ is required.\\

\begin{figure}[!htbp]
		\centering
		\includegraphics[width=0.55\linewidth]{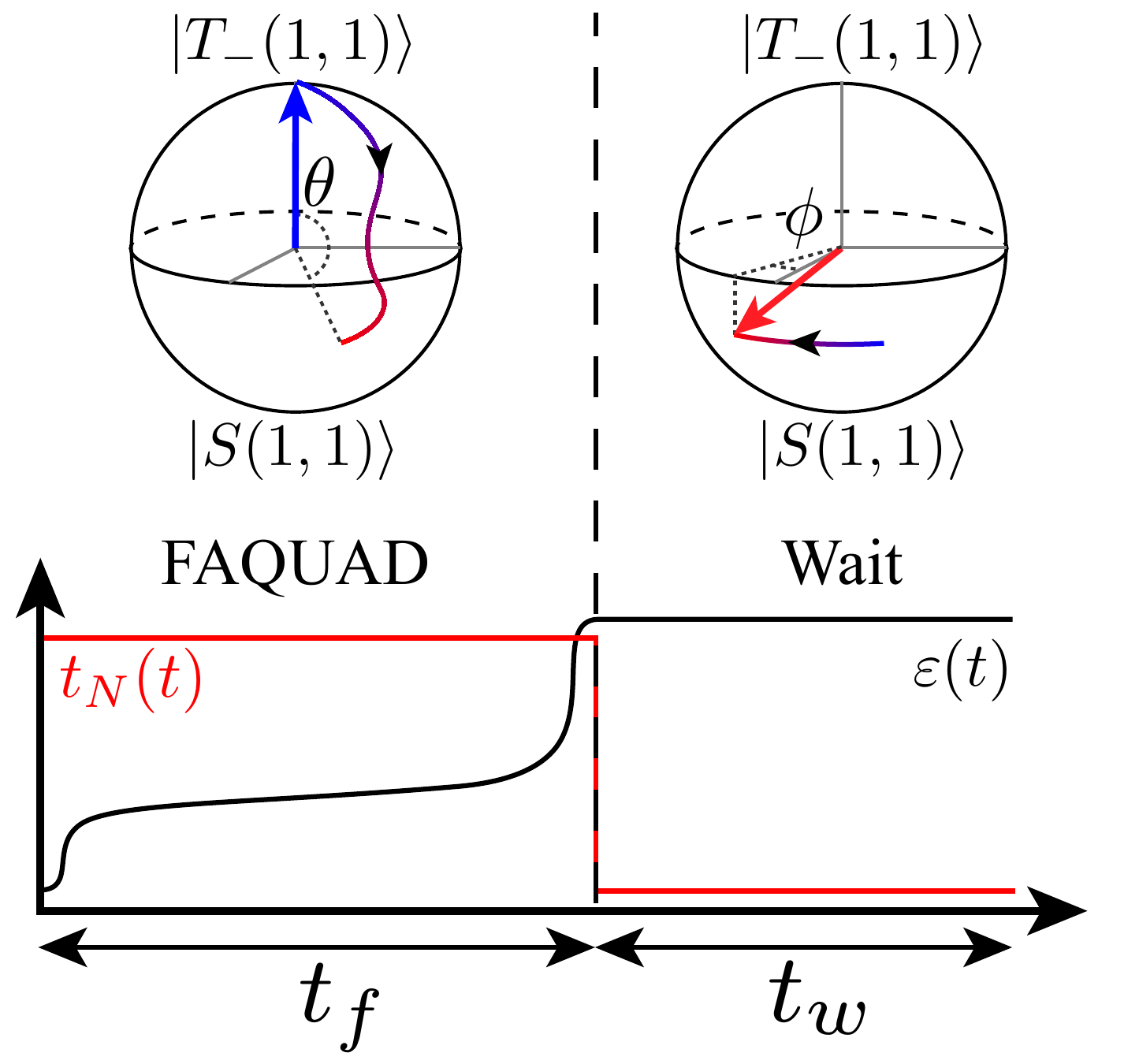}
		\caption{Scheme for the state initialization protocol. The first step consists of a FAQUAD pulse in the detuning $\varepsilon$, obtaining the final polar angle $\theta$. In the second step, the tunneling $t_N$ is set to zero, and the system evolves acquiring a phase $\phi$.}
		\label{fig:state_preparation}
	\end{figure}

	\textit{Qubit state initialization.} One of DiVincenzo's criteria for the construction of a quantum computer \cite{DiVincenzo2000} is the ability to initialize the state of the qubit. We use the singlet-triplet states to map the computational basis as $\ket{0}\equiv\ket{T_-(1, 1)}$ and $\ket{1}\equiv\ket{S(1, 1)}$. A general qubit state is written as
	\begin{equation}
		\ket{\Psi} = \cos\theta / 2 \ket{T_-(1, 1)} + e^{i\phi}\sin \theta/2\ket{S(1, 1)},
	\end{equation}
	where $\theta$ and $\phi$ are the polar and azimuthal angles, respectively, defined on the Bloch sphere. To achieve a general state, the qubit is first initialized at $\ket{T_-(1, 1)} $ using spin-selective operations on the DQD system. Another possibility of initialization is to let the system decay to its ground state, which corresponds to $\ket{T_-(1, 1)}$ in the limit $\varepsilon < \tilde{\varepsilon}$. Our goal is to develop a protocol that can be applied to evolve the system to an arbitrary value of the angles $\theta$ and $\phi$. When performing a FAQUAD protocol, we found that varying $t_f$ from $t_f\rightarrow0$ to $t_f = \tilde{t}_f$, which corresponds to the first peak in the fidelity of the state transfer, the final polar angle goes smoothly from $\theta\rightarrow 0$ to $\theta \rightarrow \pi$. Then, the polar angle can be tuned by implementing a FAQUAD pulse with a given $t_f$ that depends on the desired polar angle. The acquired azimuthal angle during the process of FAQUAD depends on the final polar angle, $\phi_\mathrm{FAQUAD}(\theta_f)$, which can be approximated by a second-order polynomial function. To achieve the desired phase, one can do a rotation around the $z$-axis by letting the system evolve in the large detuning limit. The phase during this waiting time $t_w$ reads $\phi_w = \frac{1}{\hbar}\int_0^{t_w} dt[E_{S(1, 1)} - E_{T_-(1, 1)}] = -\frac{E_z}{\hbar}t_w$. This last expression is only valid if the coupling between the states is low enough, which can be obtained using large detuning, or simply turning off the tunneling rate and setting $t_N=0$. The total phase after the waiting time is given by $\phi = \phi_\mathrm{FAQUAD}(\theta) + \phi_w(t_w)$. From here we can extract $t_w$, which depends on both $\phi$ and $\theta$. This protocol is schematically shown in Fig.~\ref{fig:state_preparation}.
	
\begin{figure}[!htbp]
		\centering
		\includegraphics[width=0.7\linewidth]{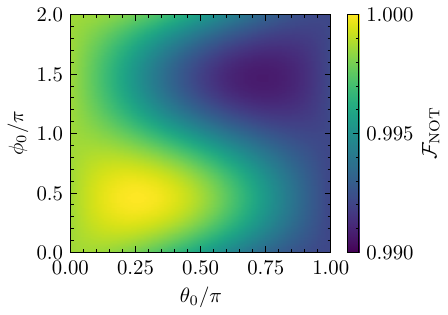}
		\caption{Fidelity of the NOT gate implemented with FAQUAD, as a function of the initial state of the qubit defined by the polar angle $\theta_0$ and the azimuthal angle $\phi_0$. $B=5$ mT, $U=2$ meV, $t_N=5\; \mu$eV, $\lambda_2=0.25\;\mu$eV, $\lambda_1=\lambda_2/100$, FAQUAD time $\tilde{t}_f=69.16$ ns, and waiting time $t_w=4.21$ ns.}
		\label{fig:NOT_gate}
	\end{figure}

	\textit{NOT Gate.} One of the main one-qubit gates in all quantum algorithms is the NOT gate (also known as $X$-gate), represented by the $\sigma_x$ Pauli matrix. The action of this gate is a $\pi$ rotation around the $x$-axis. For instance, if the qubit is initialized in $\ket{0}$, after applying the NOT gate the qubit will be in the state $\ket{1}$, and vice versa. In our system, we can implement a NOT gate by applying FAQUAD with a total time $t_f = \tilde{t}_f$ and a waiting time $t_w$ such that the dynamical phase is corrected, which is almost insensitive to the initial state. We define the fidelity of the NOT gate as $\mathcal{F}_\mathrm{NOT} = \abs{\bra{\Psi(0)}\sigma_x\ket{\Psi(t_f)}}^2$, which is shown in Fig.~\ref{fig:NOT_gate} for different initial states. The fidelity achieved for different initial polar angles ($\theta_0$) and azimuthal angles ($\phi_0$) is always $\mathcal{F}_\mathrm{NOT} > 0.99$, and the corresponding gate time is $73.37$~ns. The average fidelity for all possible initial states is $\mathcal{F}_\mathrm{NOT} = 0.995$. Using other protocols such as LA or a $\pi$-pulse lower fidelity is obtained. The results for these protocols are shown in the Supplemental Material \cite{SuppMat}.\\
	
\begin{figure}[!tp]
		\centering
		\includegraphics[width=.9\linewidth]{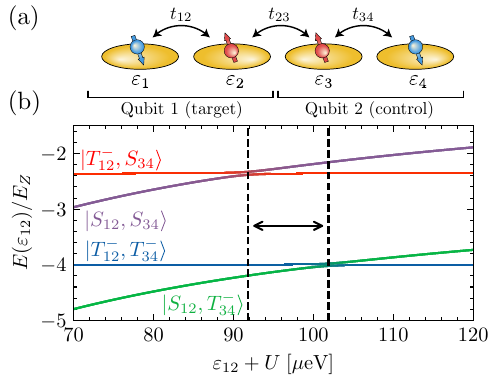}
		\caption{(a) Scheme of a quadruple QD array populated with four HH. Each pair represents one $\ket{S}$-$\ket{T_-}$ qubit. (b) Energy-level diagram versus the detuning between the two left-most quantum dots $\varepsilon_{12} \equiv \varepsilon_2 - \varepsilon_1$, while the detuning between the other dots is kept constant at $\varepsilon_{34} + U = 50\; \mu$eV. $B = 2.3$~mT, $t_{N, 12} = t_{N,34} = 10\; \mu$eV, $t_{N, 23} = 5\; \mu$eV $t_{F, ij} = 0.02 t_{N, ij}$, $U = 2$~meV.}
		\label{fig:4HH_band_structure}
	\end{figure}

	\textit{Two-qubit gates.} Now we propose to implement a CNOT gate considering a linear array of two DQDs, as shown in Fig.~\ref{fig:4HH_band_structure} (a). The system is described by the Hamiltonian $H_{2Q} = H_0^{(1)} + H_0^{(2)} + H_\text{int}$, where $H_0^{(1,2)}$ are the single qubit Hamiltonians for each DQD (see Eq.~(\ref{eq:Hamiltonian_DQD})), and $H_\text{int}=-\sum_{\sigma,\sigma'=\{\uparrow, \downarrow\}}t_{\sigma,\sigma'}\left(c_{2\sigma}^\dagger c_{3\sigma'} + h.c.\right)$ is the coupling Hamiltonian between them, with $t_{\sigma, \sigma'} = t_N$ for $\sigma=\sigma'$, and $t_{\sigma, \sigma'}=\lambda_2$ otherwise. Each qubit is defined by the singlet-triplet hole spin in the corresponding DQD. In this section, the hole spin states are labeled as $\ket{S_{ij}}$ and $\ket{T_{ij}^-}$ for the singlet and the triplet states, respectively, with one hole in each dot.
	
	One DQD defines the target qubit on which a one-qubit gate will be performed, while the other DQD is the control qubit that governs whether the operation on the target qubit will be performed. If the control qubit is in the state $\ket{0}$, no operation applies to the target qubit, in other words, the identity gate is applied leaving the target qubit in its original state. On the other hand, if the control qubit is in the state $\ket{1}$, a NOT gate is performed to the target qubit. The CNOT gate reads $U_\mathrm{CNOT} = \ket{0}\bra{0}\otimes \mathds{1} + \ket{1}\bra{1}\otimes \sigma_x$.
	
	
	
	In this case, the left-most DQD defines the target qubit, while the right-most DQD controls the two-qubit gate. The energy level diagram against detuning between the two leftmost dots is shown in Fig.~\ref{fig:4HH_band_structure} (b), while the detuning between the two right-most dots remains constant. Due to the coupling between the middle dots, the $T^-_{12}-S_{12}$ avoided anticrossing location depends on the state of the control qubit. Tuning a FAQUAD protocol to work near the transition with the control qubit in the $\ket{S_{34}}$ state ($\varepsilon_{12} + U \sim 92\; \mu$eV), the dynamics of the target qubit is reduced to the case of a single qubit, performing a NOT gate. However, using the same pulse with the control qubit in the $\ket{T_{34}}$ state, there is no anticrossing in the working detuning regime, so the target qubit remains in the same initial state since the dynamic is diabatic. It is the shift in energies between both anticrossings due to the coupling between the second and the third QDs that makes this protocol possible. The fidelity of the quantum gate will increase with the difference in detuning for each anticrossing. Furthermore, we can also tune the FAQUAD protocol to work near the other anticrossing ($\varepsilon_{12} + U \sim 102\; \mu$eV), resulting in a quantum gate given by a NOT gate over the control qubit, a CNOT gate, and finally another NOT gate over the control qubit.
	
	By tuning the total time of the protocol and the magnetic field intensity, we have control over the acquired phases. Setting the total time of the FAQUAD protocol to $t_f = 182.73$ ns, and a waiting time of $t_w=5.6$~ns, the gate applied corresponds to a pure CNOT. The explicit value of the unitary transformation at the given total time is
	\begin{widetext}
		\begin{equation}
			U_\mathrm{CNOT}(t_f) = \bordermatrix{~ & \ket{0, 0} & \ket{1, 0} & \ket{0, 1} & \ket{1, 1}\cr
				~ & 1 & 0.05+0.02i & 0.01 & 0.02 \cr
				~ & -0.05+0.02i & 0.99 & -0.05+0.01i & 0.05-0.01i \cr
				~ & -0.01 & -0.05 & -0.03 & 0.99-0.12i \cr
				~ & -0.01 & 0.05+0.01i & 0.99+0.08i & 0.03}\; .
		\end{equation}
	\end{widetext}
	To compute the fidelity of the two-qubit gate we use \cite{Song_2022,Pedersen2007}
	\begin{equation}
		\mathcal{F}_\text{2Q} = \frac{1}{d(d+1)} \left[\Tr\left(MM^\dagger\right) + \abs{\Tr(M)}^2\right],
	\end{equation}
	where $d=4$ is the dimension of the computational space and $M\equiv U_\text{tar}^\dagger U(t_f)$ with $U_\text{tar}$ the target unitary evolution matrix, i.e., the NOT gate defined above. Using our proposed protocol we obtain a CNOT gate fidelity of $\mathcal{F}_\text{CNOT} = 0.99$. This two-qubit gate fidelity is comparable with other proposals which use electrons instead of holes (see Ref.~\cite{Ribeiro_2010}).
	
\begin{figure}[h!]
		\centering
		\includegraphics[width=0.9\linewidth]{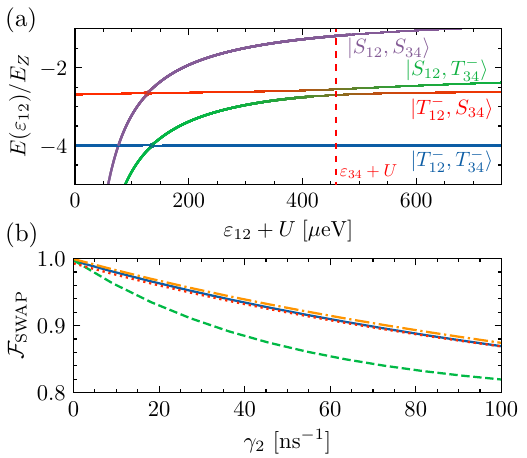}
		\caption{(a) Energy-level diagram versus the detuning between the two left-most QDs $\varepsilon_{12} \equiv \varepsilon_2 - \varepsilon_1$. The colors represent the population of the basis states in each instantaneous eigenstate. (b) Fidelity of the SWAP gate obtained with FAQUAD (solid blue), linear (dashed green), $\pi$-pulse (dotted dashed orange), and LA (dotted red) as a function of charge noise. $\varepsilon_{34} + U = 459\; \mu$eV, $B = 5$ mT, $t_{N, 12} = t_{N,34}= 10\; \mu$eV, $t_{N,23}=4.8\; \mu$eV, $\lambda_{2, ij} = 0.02 t_{N, ij}$, $U = 2$ meV. The protocols times are $(38, 69, 37, 39.6)$~ns and the waiting times are $(2.55, 2.46, 0.44, 1.16)$~ns for FAQUAD, linear, $\pi$-pulse and LA, respectively.}
		\label{fig:4HH_band_structure_SWAP}
	\end{figure}

	For higher detuning values, near $\varepsilon_{12} \sim \varepsilon_{34}$ there is an anticrossing between $\ket{S_{12}, T_{34}^-}$ and $\ket{T_{12}^-,S_{34}}$, see Fig.~\ref{fig:4HH_band_structure_SWAP} (b). By applying FAQUAD at this avoided anticrossing we can achieve a SWAP-like gate. A pure SWAP gate is a two-qubit gate that interchanges the states of two qubits. We refer to SWAP-like when non-zero phases are allowed for non-diagonal elements. SWAP gates are also one of the most used two qubits gates for implementing quantum algorithms since it allows to couple distant qubits by sequentially transferring the quantum information between neighboring dots
	\begin{equation}
		U_\mathrm{SWAP-like} = \mqty(
		1 & 0 & 0 & 0 \cr
		0 & 0 & e^{\varphi_1} & 0 \cr
		0 & e^{\varphi_2} & 0 & 0 \cr
		0 & 0 & 0 & 1)\; .
		\label{eq:Unitary_SWAP}
	\end{equation}
	
	Using a FAQUAD pulse during a total of $t_f=38$ ns, we obtain the following evolution matrix
	\begin{widetext}
		\begin{equation}
			U_\mathrm{SWAP}(t_f) = \mqty(1 & -0.015-0.02i & -0.004+0.002i & 0 \\
			0.003-0.005i & 0.061-0.051i & -0.011-0.996i & -0.008+0.011i \\\
			0.015-0.001i & 0.993+0.082i & -0.055+0.057i & -0.007 \\\
			0 & 0.008 & 0.011-0.008i & 0.999)\; .
		\end{equation}
	\end{widetext}
	
	We can see that the obtained two-qubit gate corresponds to a SWAP gate with some additional phases in the off-diagonal matrix elements. However, a pure SWAP gate can be recovered with two additional local gates on each qubit. The fidelity obtained with FAQUAD is $\mathcal{F}_\mathrm{SWAP}=0.995$, beyond the error correction threshold. In Fig.~\ref{fig:4HH_band_structure_SWAP} (b) we compare the fidelity of the SWAP gate for different protocols against charge noise. This noise source is modeled as a pure dephasing mechanism whose strength is proportional to the population of the double-occupied state. Here the average gate fidelity is defined using the Haar measure over the quantum states of two qubits \cite{Nielsen2010}. We find that the linear ramp is very sensitive to charge noise, since the total time needed for the gate $t_f\sim 69$~ns is much larger than for the other three protocols, FAQUAD, LA, and a $\pi$-pulse, with $t_f \sim 38$~ns. A detailed discussion on the effect of systematic and stochastic errors on $\varepsilon_{12}$ and $t_{N, 12}$ can be found in \cite{SuppMat}.
	
	The parameters used in this section are selected such that high-fidelity quantum gates are obtained. However, a more exhaustive analysis can be performed to find even higher fidelities. This detailed exploration can be obtained with the help of numerical methods such as gradient descent algorithms which explore the multidimensional space spanned by the different parameters of the system, e.g., the tunneling rates, the magnetic field intensity, or the detuning between dots 3 and 4. However, this analysis is beyond the scope of this work. \\
	
	\textit{Experimental implementation.} During our work, we have focused on the study of GaAs QDs. However, our analysis can be extended to other semiconducting materials, such as Si or Ge, where the main features of FAQUAD are still valid. It can be also implemented in materials that present an inhomogeneous g-factor (see Supp. Material \cite{SuppMat}).\\
	
	\textit{Conclusions.} In this work, we propose a fast quasi-adiabatic protocol to implement the S-T hole spin qubit transition in a DQD. As the dynamic follows the adiabatic trajectory in a fast way, we are able to highly decrease charge noise by reducing the population of the double-occupied singlet state as compared with other protocols. The reduced total time of the protocol also makes it robust against spin dephasing. By means of this all-electrical protocol, we can initialize the qubit in an arbitrary state with high fidelity. Furthermore, we are able to perform a single-qubit gate, a NOT gate, combining the FAQUAD pulse and a waiting period to account for additional phases. We extend our scheme to two DQDs, each DQD representing one qubit. Driving one of the qubits allows for a two-qubit gate, a CNOT gate with $0.99$ of fidelity. We also propose how to implement a two-qubit SWAP-like gate, achieving a fidelity of $0.995$. In both cases, the obtained fidelity is beyond the fault-tolerance error correction threshold. Finally, we compare the fidelity of different experimentally used protocols against the charge noise strength.
	
	Our results demonstrate the feasibility of the proposed protocol for the implementation of one- and two-hole spin qubit gates with high fidelity. These results are a step towards the logic gates implementation with hole spin qubits in semiconductor quantum dots.\\
	
	\textit{Acknowledgments}
	We acknowledge Dr. M. Benito, Prof. E. Sherman, and L. Oltra for insightful discussions and critical reading of the manuscript. G.P. and D.F.F. were supported by Spain's MINECO through Grant No. PID2020-117787GB-I00 and by CSIC Research Platform PTI-001. D.F.F. acknowledges support from the FPU program No. FPU20/04762. Y.B. acknowledges the EU FET Open Grant Quromorphic (828826) and the QUANTEK project (ELKARTEK program from the Basque Government, expedient no. KK-2021/00070).
	
	\clearpage
	\newpage
	
	\begin{widetext}
		
		\setcounter{equation}{0}
		\setcounter{figure}{0}
		\setcounter{section}{0}
		\makeatletter
		
		\renewcommand{\thefigure}{SM\arabic{figure}}
		\renewcommand{\thesection}{SM\arabic{section}}  
		\renewcommand{\theequation}{SM\arabic{equation}} 

		\section{Spin-orbit coupling model}
		Following Ref.~\cite{Bogan2018}, we assume that the orbitals are centered at each dot and can be expressed by Gaussian waves as
		
		\begin{eqnarray}
			\psi_L &=& \frac 1l \sqrt{\frac 2\pi}\exp(-\frac{x^2 + y^2}{l^2}), \\
			\psi_R &=& \frac 1l \sqrt{\frac 2\pi}\exp[-\frac{(x - d)^2 + y^2}{l^2}],
		\end{eqnarray}
		where $l$ is the extent of the wave function and $d$ is the distance between the dots on the $x$ axis. Using the wave functions defined above, we can write the tunneling matrix element as
		\begin{equation}
			\begin{split}
				\bra{\uparrow, 0} H_\text{SOC}\ket{0, \downarrow} &=  -\hbar^3\alpha E_\perp\frac{d^3}{l^6}\exp(-\frac{d^2}{2l^2}) \\
				& \;-i\beta\hbar^3\left(\frac{4d}{l^4} - \frac{d^3}{l^6} + \frac{d}{4l_B^4}\right) \times \\
				& \;\exp(-\frac{d^2}{2l^2})\\
				&\equiv \lambda_2,    
			\end{split}
		\end{equation}
		where we have introduced the magnetic length $l_B = \sqrt{\frac{\hbar}{m\omega_c}}$, with $\omega_c = \frac{eB}{mc}$ the cyclotron frequency. For the parameters used in this work $l_B \gg d$, so we can neglect the term proportional to $d/2l_B^4$. Note that the matrix element is complex, then $\lambda_2 = \abs{\lambda_2}e^{i\vartheta}$. We find that the spin-flip phase does not play an important role in the main results of the work, so we set $\vartheta = 0$. Also, we assume that both spin-flip and spin-conserving tunneling terms are proportional to each other $\lambda_2 = x_\text{SOC}t_N$, where $x_\text{SOC}$ is a constant \cite{Stepanenko_2012,Mutter2021}. The value of $x_\text{SOC}$ is given by the experimental device fabrication parameters, such as the distance between dots. In the presence of Rashba SO interaction, the parameter $x_{\mathrm{SOC}}$ can also be modified by electric fields.
		
		\section{Local Adiabatic scheme}
		A possible modification for FAQUAD is to assume that $\bra{\phi_i(\varepsilon)}\partial_\varepsilon H \ket{\phi_k(\varepsilon)} = 1$ for all values of $k\neq i$. With this assumption, we can write Eq. (7) of the main text for the driving parameter as
		\begin{equation}
			\dot{\varepsilon} = \frac{c}{\hbar}\sum_{k\neq i} \abs{E_i(\varepsilon) - E_k(\varepsilon)}^\alpha
		\end{equation}
		with $\alpha = 2$. Then, the driving parameter has only information about the energy levels of the system and does not need knowledge of the eigenstates. This modification of the FAQUAD protocol is known as Local Adiabatic (LA) \cite{Roland_2002,Richerme_2013}. A generalized version of LA can be obtained by taking into account different values of the exponent $\alpha\neq2$. In Fig.~\ref{fig:LA_exponent} (a) we show how the detuning pulse shapes depend on the exponent $\alpha$. For $\alpha=0$ a linear ramp is recovered. As the exponent grows, the pulse develops an abrupt change close to the start and end of the protocol, while for intermediate times the pulse is flatter. This behavior ensures adiabatic dynamics, but it can be detrimental to a possible experimental implementation due to the sharp changes in the detuning. In Fig.~\ref{fig:LA_exponent} (b) we show how the first peak in fidelity ($\tilde{\mathcal{F}}$) and the total time needed to reach this fidelity ($\tilde{t}_f)$ depend on the exponent value. For the values $\alpha < 0.5$, there is no peak in the fidelity for times $t_f < 300$~ns. Between $0.5<\alpha < 1.4$ the fidelity of the protocol increases up to values close to one, while the total time needed is relatively high compared with FAQUAD ($\tilde{t}_f^{\;\mathrm{FAQUAD}}\sim 100$ ns). For values of $\alpha > 1.4$, the fidelity develops a maximum for shorter times, as can be seen in the abrupt change in the total time $\tilde{t}_f$ that is needed to reach the peak in fidelity. This change in total time is followed by a small decrease in the fidelity, which rapidly recovers high values for $\alpha\sim2$. Since the pulses for large values of $\alpha$ are similar to a $\pi$-pulse, we predict that the minimum time needed to reach a high fidelity is obtained in this regime. This is indeed what happens and the minimum total time needed to reach the first peak decreases up to values close to $\tilde{t}_f\sim 60$~ns. The first peak of fidelity is reached with $\alpha=2.18$, which is the value chosen to compare LA with the other pulses in Fig.~4 of the main text.
		\begin{figure}[!htbp]
			\centering
			\includegraphics[width=0.8\linewidth]{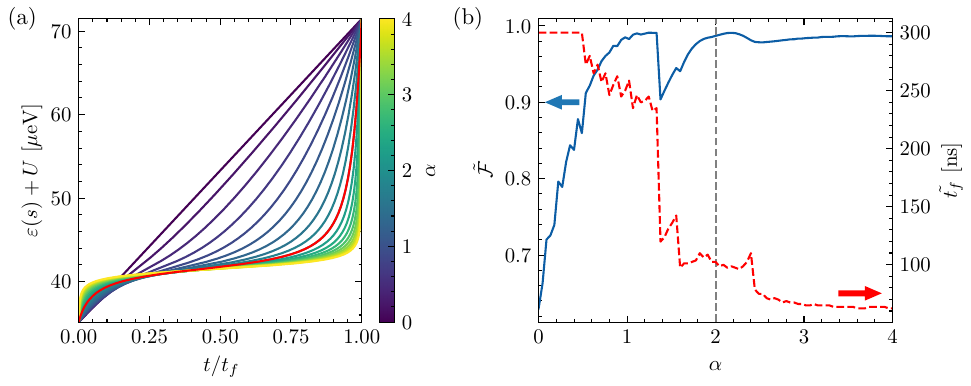}
			\caption{(a) Detuning pulse shapes (color coded) depending on the local adiabatic exponent $\alpha$. The original $\alpha=2$ is shown with a red line. (b) First peak on fidelity  $\tilde{\mathcal{F}}$ (left axis, blue solid line) and total time needed to reach this fidelity $\tilde{t}_f$ (right axis, red dotted-dashed line) for a LA protocol with exponent $\alpha$. The gray dashed line denotes the original LA proposal with $\alpha=2$. $B = 15$ mT, $U=2$~meV, $t_N=5\; \mu$eV, $\lambda_2=0.1\; \mu$eV, $\lambda_1 = \lambda_2 / 100$.}
			\label{fig:LA_exponent}
		\end{figure}
		
		\section{LZ-type protocol}
		The original Landau-Zener formula for transfer through a two-level system was derived using a single linear pulse that drives the system between the avoided anticrossing. The probability for the transfer, in this case, is given by the Landau-Zener formula $\mathcal{F} = 1 - \exp(-2\pi a^2 / \hbar v)$, where $a$ is the coupling between the two states, and $v$ the slope of the linear ramp. However, as we show in Fig.~4 of the main text, the time needed to achieve high fidelity is quite large compared to other conventional protocols such as a $\pi$-pulse. In order to improve the simple linear ramp, one can use a three-step pulse consisting of two linear ramps with a total raise time given by $t_\text{raise}$ for each one, and a waiting time in between $t_\text{wait}$ where the detuning is kept constant. During the first step, we drive the system through the avoided anticrossing. At the end of the finite ramp time, there is an admixture between the two states. The waiting time allows each state to acquire a different phase due to the energy gap between levels. This relative phase is responsible for constructive and destructive interferences during the returning pulse, suppressing the total population for one of the two states. By tuning the ramp and waiting times, we can obtain results that improve the fidelity as compared with simple linear ramp protocols. In Fig.~\ref{fig:LZS_protocol} we study the fidelity reached with this three-step protocol for a state transfer in terms of the raise and the waiting time. The oscillations with a fixed value of $t_\mathrm{raise}$ are due to the periodic relative phase obtained during the waiting time. We find that there are fringes with fidelities close to $\mathcal{F} \sim 0.9$, but lower than those obtained with FAQUAD (see Fig. 4 of the main text) in this range of times.
		\begin{figure}[!htbp]
			\centering
			\includegraphics[width=0.45\linewidth]{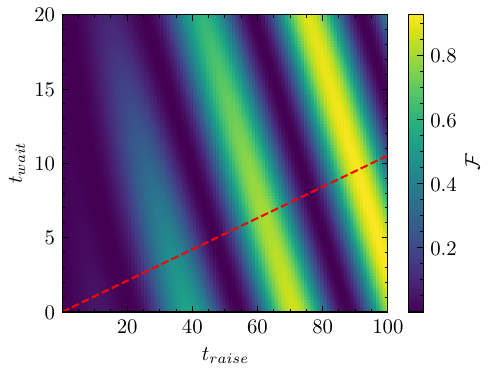}
			\caption{Dependence of the fidelity for a state transfer on the raise time and the waiting time for a LZ protocol. The dashed red line denotes the fidelity shown in Fig.~4 of the main text, for $t_\text{wait} = 0.1t_\text{raise}$. $B = 10$ mT, $U=2$ meV, $t_N=5\; \mu$eV, $\lambda_2=0.1\; \mu$eV, $\lambda_1 = \lambda_2 / 100$.}
			\label{fig:LZS_protocol}
		\end{figure}
		
		\section{Inhomogeneous \texorpdfstring{$g$}{g}-factor} \label{sec:anisotropic}
		
		Recently, it has been shown that due to the quantum dot geometry, holes in Ge have an inhomogeneous site-dependent g-tensor \cite{Hofmann_2019,Mutter2021, Mutter2021b, Jirovec2022}. In this section, we will study the effect of this contribution on our transfer protocols. A general Hamiltonian for the system can be written as
		\begin{equation}
			H_T = H_0 + H_Z + H_\text{tunneling},
		\end{equation}
		where the first term corresponds to the individual dot energy $\varepsilon_i$ and the intradot Coulomb repulsion
		\begin{equation}
			H_0 = \sum_{i=1}^{N}\varepsilon_i \hat{n}_i + U_i\hat{n}_{i\uparrow}\hat{n}_{i\downarrow},
		\end{equation}
		with $N$ the total number of dots in the system. The term owing to a constant magnetic field $\vec{B} = (B_x, B_y, B_z)$ and a site-dependent $g_i$ tensor is
		\begin{equation}
			H_Z = \sum_{i=1}^N (\vec{B}\cdot g_i)\cdot \vec{\hat{\sigma}}_i,
		\end{equation}
		with $\vec{\hat{\sigma}}_i$ the Pauli matrices acting on the $i$ site. In order to describe the tunneling Hamiltonian, which is a combination of the spin-conserving tunneling rate $t_N$ and the spin-flip tunneling rate due to the spin-orbit coupling $t_{\text{SO}} = (t_x, t_y, t_z)$, we define the site-dependent matrix
		\begin{equation}
			\mathbf{t}^n \equiv t_N ^n \mathds{1}_{2\times 2} -i \sum_{\alpha=\{x, y, z\}}t_\alpha^n\cdot \sigma_\alpha = \begin{pmatrix}
				t_N^n -it_z^n & -it_x^n - t_y^n \\
				-it_x^n + t_y^n & t_N^n +it_z^n 
			\end{pmatrix},
		\end{equation}
		where $t^n_{(N, x, y, z)}$ are the different tunneling rates between two adjacent dots. With this matrix, we can write all the tunneling elements in the Hamiltonian in a compact way as
		\begin{equation}
			H_\text{tunneling} = \sum_{i=1}^{N-1} \sum_{\sigma,\sigma^\prime = \{\uparrow, \downarrow\} } t_{\sigma,\sigma^\prime}^i\left(\hat{c}_{i,\sigma}^\dagger\hat{c}_{i + 1,\sigma^\prime} + \hat{c}_{i + 1,\sigma}^\dagger\hat{c}_{i,\sigma^\prime}\right).
		\end{equation}
		For a DQD the super index of the tunneling rates can be suppressed. Working on the molecular basis spanned by the triplets $\{|T_\pm\rangle, |T_0\rangle \}$ and singlet states $\{|S(1, 1)\rangle, |S(0, 2)\rangle \}$ the total Hamiltonian is written as \cite{Mutter2021b}
		\begin{equation}
			\begin{split}
				H_T = & (\varepsilon + U) \ketbra{S(0, 2)} + \sqrt{2}t_N\ket{S(1, 1)}\bra{S(0, 2)}\\
				& + i\sqrt{2}t_z\ket{T_0}\bra{S(0, 2)} - \sum_{\pm}(t_y \pm it_x)\ket{T_\pm}\bra{S(0, 2)}\\
				& + B_z\left(\sum_\pm g_z^\pm \ketbra{T_\pm} + g_z^-\ket{S(1, 1)}\bra{T_0}\right)\\
				& +\sum_\pm \frac{B_x \pm iB_y}{\sqrt{2}}\left(g_x^+\ket{T_0}\bra{T_\pm}\mp g_x^-\ket{S(1, 1)}\bra{T_\pm}\right)\\
				& + \text{H.c.},
			\end{split}
		\end{equation}
		where we have assumed a diagonal $g$ tensor in the form $g = \operatorname{diag}(g_x, g_x, g_z)$, and defined the quantities $g_\alpha^\pm = g_\alpha^L \pm g_\alpha^R$, where $L$ ($R$) refers to the left (right) dot. To simplify the model, we consider a magnetic field perpendicular to the quantum dot plane $\vec{B} = (0, 0, B)$. We also assume a spin-orbit tunneling vector parallel to the QD plane as $t_\text{SO} = (\lambda_2, \lambda_2, 0) /\sqrt{2}$, with $|t_\text{SO}| = \lambda_2$.
		
		The protocols based on adiabatic (or quasi-adiabatic) dynamics need an instantaneous eigenstate that connects the desired initial and final states. In Fig.~\ref{fig:adiabatic_states} (a), we plot the population of the different basis elements in a given instantaneous eigenstate, $\ket{\phi}=c_1(\varepsilon)\ket{S(1, 1)} + c_2(\varepsilon)\ket{S(0, 2)} + c_3(\varepsilon)\ket{T_-}$, that connects $\ket{T_-}$ (initial state) in the low detuning regime with $\ket{S(1, 1)}$ (target state) in the high detuning regime. These populations are computed in the case in which both dots have the same $g$-factor, i.e., $g_z^-= 0$. However, if the material of interest has an inhomogeneous $g$-factor, as in the case of Ge, the adiabatic state no longer fully populates the target state $\ket{S(1, 1)}$, see Fig.~\ref{fig:adiabatic_states} (b). In fact, the instantaneous eigenstate in the large $\varepsilon$ limit is given by $(\ket{T_0} - \ket{S(1, 1)}) / \sqrt{2}$.
		\begin{figure}[!htbp]
			\centering
			\includegraphics[width=0.7\linewidth]{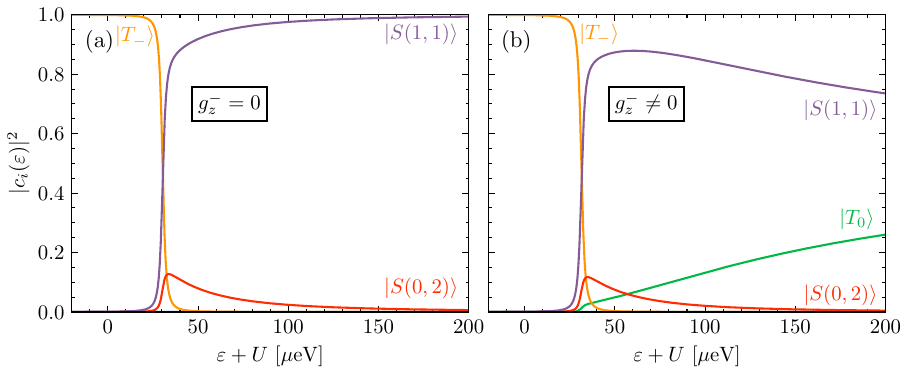}
			\caption{Population $\abs{c_i(\varepsilon)}^2$ of the different basis elements in the instantaneous adiabatic state of interest $\ket{\phi}$, versus the detuning for (a) constant g-factor for all dots, i.e., $g_z^L = g_z^R$, and (b) inhomogeneous g-factor with $g_z^- = 2.04$. In both cases, $g_z^+ = 12$, $B = 20$~mT, $t_N = 11.38\; \mu$eV, $\lambda_2 = 0.392 \; \mu$eV, and $U = 2$~meV \cite{Jirovec2022}.}
			\label{fig:adiabatic_states}
		\end{figure}
		
		The fidelity of FAQUAD is then sensitive to the coupling between $\ket{T_0}$ and $\ket{S(1,1)}$ due to the inhomogeneous g-factor. In Fig.~\ref{fig:fidelity_g_minus} we plot the maximum population for the single-occupied singlet state in the instantaneous eigenstate against $g_z^-$. As expected, the maximum population decreases with the inhomogeneous g-factor, as does the first peak in the fidelity for the $\ket{T_-}\rightarrow \ket{S(1, 1)}$ transfer using FAQUAD. One possible solution to recover high-fidelity transfer in materials with large g inhomogeneity is by means of a magnetic field gradient. If the Zeeman splitting is the same for all dots $E_Z^{L} = E_Z^R$, that is, $B^Lg_z^L = B^Rg_z^R$, the states $\ket{T_0}$ and $\ket{S(1, 1)}$ are effectively uncoupled, recovering the high-fidelity transfer discussed in the main text.
		\begin{figure}[!htbp]
			\centering
			\includegraphics[width=0.4\linewidth]{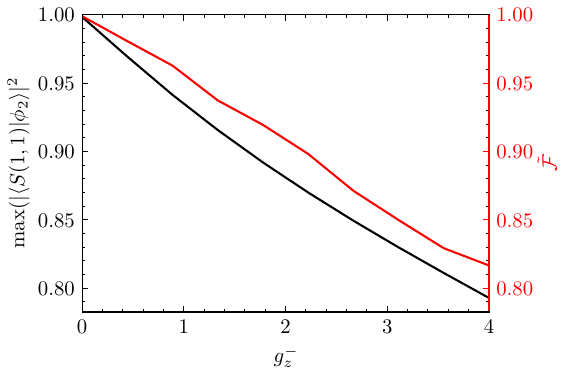}
			\caption{(Left axis, black solid line) The maximum population of $S(1, 1)$ in the instantaneous eigenstate of interest $\ket{\phi}$, and (right axis, red solid line) first peak in the fidelity using FAQUAD, versus the inhomogeneous g-factor in the $z$-axis. The other parameters are the same as those in Fig~\ref{fig:adiabatic_states}.}
			\label{fig:fidelity_g_minus}
		\end{figure}
		
		Another possible way to achieve high-fidelity transfer with non-zero $g$-factor inhomogeneity is to change the driving parameter. In this case, the detuning is kept constant. The driving parameter will be now the magnetic field, recovering an adiabatic state that connects $\ket{T_-}$ with $\ket{S(1,1)}$ in the limits of a low and high magnetic field, respectively (see Fig.~\ref{fig:FAQUAD_B_all} (a), blue line). In Fig.~\ref{fig:FAQUAD_B_all} (b), we plot the fidelity for the transfer using the magnetic field as the driving parameter, obtaining values close to one in short total times. In the inset, we can see that the maximum fidelity obtained with a FAQUAD pulse increases as we move further away from the avoided crossing between hybridized singlet states, located at $\varepsilon = -U$. The same protocol can also be employed in the case of uniform g-factor $g_z^- = 0$, obtaining similar results. However, in the main text, we opted to focus on detuning as the driving parameter, since it is more feasible in an experimental set-up, rather than a time-dependent magnetic field.
		
		\begin{figure}[!htbp]
			\centering
			\includegraphics[width=0.8\linewidth]{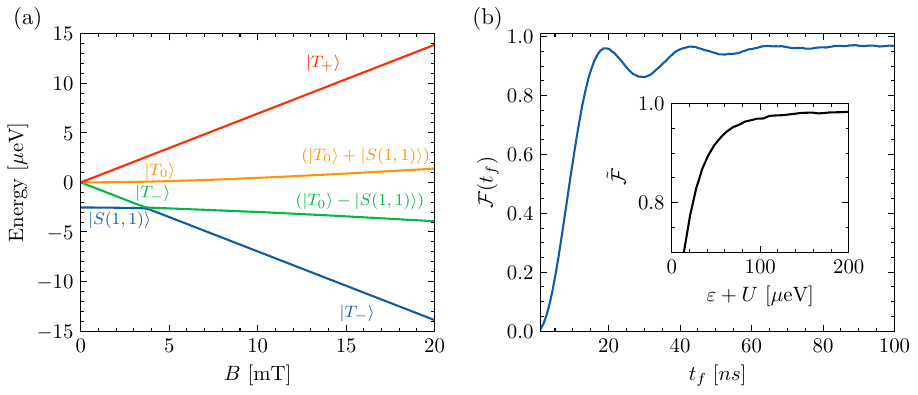}
			\caption{(a) Energy-level diagram for two HH in a DQD versus the magnetic field. The detuning between dots is $\varepsilon = -U + 100\; \mu$eV. The state $\ket{S(0, 2)}$ is high in energy and is not shown here. (b) Fidelity for the $\ket{T_-} \rightarrow \ket{S(1, 1)}$ transfer versus the total time of the protocol. (Inset) Maximum fidelity for the transfer against the detuning between dots. The other parameters are the same as those in Fig.~\ref{fig:adiabatic_states}.}
			\label{fig:FAQUAD_B_all}
		\end{figure}
		
		\section{NOT gate fidelity}
		In this section we show the dependence of the NOT gate fidelity versus the initial state given by $\ket{\Psi(0)} = \cos(\theta_0 / 2) \ket{0} + e^{i\phi_0}\sin(\theta_0/2)\ket{1}$. Note that we have used the computational basis, defined as $\ket{0}\equiv \ket{T_-(1,1)}$ and $\ket{1}\equiv \ket{S(1, 1)}$. The fidelity of the gate is computed as $\mathcal{F}_\mathrm{NOT} \equiv \abs{\braket{\Psi(t_f)}{\Psi_T}}^2$, where $\ket{\Psi_T} = X\ket{\Psi(0)} = e^{i\phi_0}\sin(\theta_0 / 2) \ket{0} + \cos(\theta_0/2)\ket{1}$ is the target final state, and $X$ the Pauli matrix. In Fig.~\ref{fig:NOT_protocols}, we plot these results for three protocols: FAQUAD, LA, and a $\pi$-pulse. The best results are obtained with FAQUAD, resulting in an average fidelity of $\mathcal{F}_\mathrm{NOT} = 0.995$, followed by the $\pi$-pulse with $\mathcal{F}_\mathrm{NOT} = 0.993$, and the LA protocol with $\mathcal{F}_\mathrm{NOT} = 0.992$. Furthermore, FAQUAD obtains a broader high-fidelity range of initial states (yellow region), compared with the $\pi$-pulse, which obtains large fidelity in a very narrow region.
		
		\begin{figure}[!htbp]
			\centering
			\includegraphics[width=0.9\linewidth]{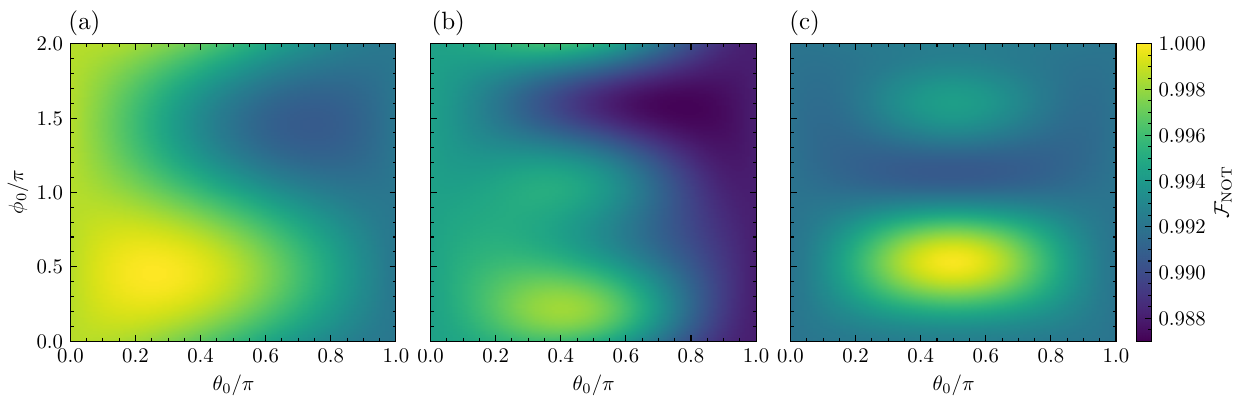}
			\caption{Fidelity for the NOT gate implemented with (a) FAQUAD, (b) LA, (c) $\pi$-pulse for different initial states of the system defined by the polar angle $\theta_0$ and the azimuthal angle $\phi_0$. $B = 5$ mT, $U=2$ meV, $t_N=5\; \mu$eV, $\lambda_2=0.25\; \mu$eV, $\lambda_1 = \lambda_2 / 100$, protocol times $t_f^{\text{FAQUAD}} = 69.16$ ns, $t_f^{\text{LA}} = 120.2$ ns, $t_f^{\pi\text{-pulse}} = 64$ ns, and waiting times  $t_w^{\text{FAQUAD}} = 4.21$ ns, $t_w^{\text{LA}} = 1.8$ ns, $t_w^{\pi\text{-pulse}} = 0$ ns.}
			\label{fig:NOT_protocols}
		\end{figure}	
		
		\section{Error in SWAP-like gate}
		To complete the section of the two-qubit gates in the main text, we analyze different noise sources and how they affect the fidelity of the proposed gate. First, we study the effect of a systematic error in some of the parameters of the system, e.g., the detuning between the leftmost two dots or the tunneling matrix elements between these QDs. A constant shift in the detuning can be written as $\varepsilon_{12} \rightarrow \varepsilon_{12}(1 + \delta_{\varepsilon_{12}})$. Note that $\varepsilon_{12}$ is the driving parameter in FAQUAD, while other QDs on-site energies are kept constant. In a similar way, a systematic error in the tunneling between the first two dots is given by $t_{N, 12} \rightarrow t_{N,12}(1+\delta_{t_{12}})$. This systematic error is also applied to the spin-flip term $\lambda_{2} \rightarrow \lambda_{2}(1+\delta_{t_{12}})$.
		
		\begin{figure}[!htbp]
			\centering
			\includegraphics[width=0.8\linewidth]{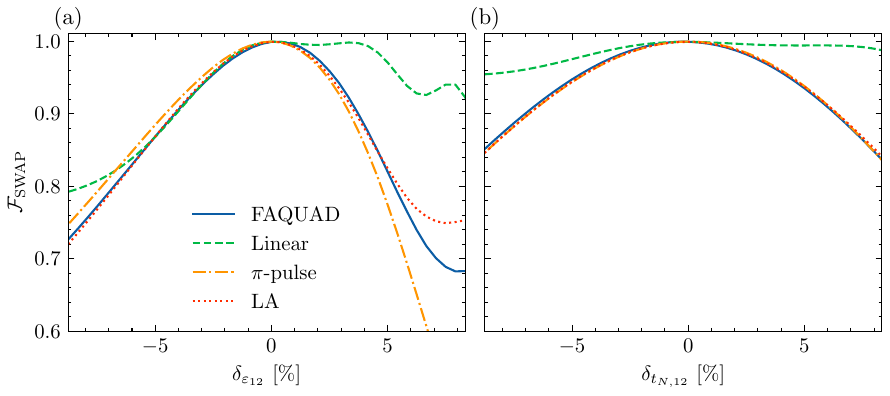}
			\caption{SWAP fidelity against systematic error in (a) the detuning, and (b) the tunneling rate between the first two QDs. The gate is implemented with FAQUAD (blue solid lines) with a total time of $t_f = 38$ ns, a linear pulse (green dashed lines) with a total time of $t_f = 69$ ns, a $\pi$-pulse (orange dotted-dashed lines) with $t_f = 37$ ns of duration, a LA (red dots) with $t_f = 39.6$ ns. The other parameters are the same as those in Fig.~9 of the main text.}
			\label{fig:SWAP_systematic}
		\end{figure}
		
		The fidelity is expected to be highly sensitive to a systematic shift in the detuning if the gate is implemented with FAQUAD, which is susceptible to the exact location of the anticrossing. This location also depends on the tunneling rate between QDs 1 and 2, so the fidelity of the SWAP gate is sensitive to changes in this parameter. However, in both cases, the fidelity reaches values greater than $0.99$ for $|\delta| < 1\%$, see Fig.~\ref{fig:SWAP_systematic} (blue solid lines). For comparison, we include the fidelity of the SWAP gate implemented with a linear pulse in the detuning (green dashed lines). The linear ramp has no information on the energy level diagram of the system so it is very robust against a systematic error in the parameters. However, as we will see below for other different sources of noise, a linear pulse gives worse results than FAQUAD. We found that concerning systematic noise in both the detuning and the tunneling rate, the results obtained with the $\pi$-pulse, LA and FAQUAD are quite similar. All these protocols need precise information on the energy-level diagram, and a constant change in some of the parameters of the system leads to low-fidelity protocols.\\

		Another possible source of error is a stochastic fluctuation in the value of the parameters. Assuming zero-mean amplitude and uncorrelated at different times noise, the effect can be described with the master equation \cite{Ruschhaupt2012}
		\begin{equation}
			\frac{d}{dt}\rho = -\frac{i}{\hbar}[H_0,\rho] - \frac{\gamma^2}{\hbar^2}\sum_i[H_i,[H_i,\rho]],
		\end{equation}
		where $H_i$ is the driving Hamiltonian. In our case, we consider an independent noise for the on-site energies of each of the first two QDs, i.e., $H_1(t) = \varepsilon_1(t) c_1^\dagger c_1$ and $H_2(t) = \varepsilon_2(t) c_2^\dagger c_2$. The noise amplitude associated to the fluctuation in the detuning is given by $\gamma_{\varepsilon_12}$. In experimental set-ups there exist cross-talk between the different plungers, and is reasonable to assume that the stochastic noise also affects the tunneling elements. Therefore, we also include the term $H_t = t_{N, 12}\sum_\sigma\left( c_{1\sigma}^\dagger c_{2\sigma} + h.c.\right) + \lambda_2\sum_{\sigma\neq \sigma'}\left(c_{1\sigma}^\dagger c_{2\sigma'} + h.c.\right)$, with a noise strength of $\gamma_{t_{N, 12}}$. In Fig.~\ref{fig:SWAP_stochastic} we show the results obtained by solving the master equation for different protocols. Similar to a systematic error, the fidelity of all the studied protocols is more robust against a stochastic error in the tunneling parameters than in the detuning. As we mentioned before, the linear ramp is more sensitive to stochastic noise, both in the detuning and tunneling matrix elements, than the other protocols. A possible explanation for this behavior is that the linear ramp takes longer to achieve a high-fidelity SWAP gate, so the pulse is exposed to stochastic noise for more time than the FAQUAD pulse. Once again, FAQUAD, a $\pi$-pulse, and LA obtain similar results against stochastic noise in the driving parameter and the tunneling rate.
		
		\begin{figure}[!htbp]
			\centering
			\includegraphics[width=0.8\linewidth]{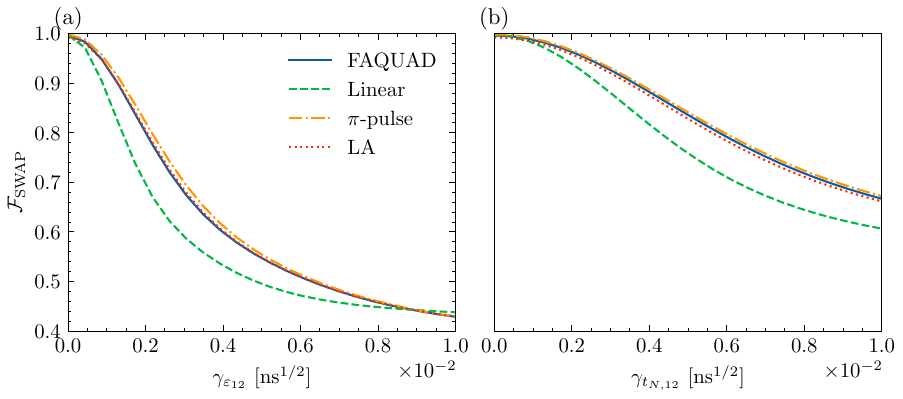}
			\caption{SWAP fidelity against stochastic error in (a) the detuning, and (b) the tunneling rate between the first two QDs. The gate is implemented with FAQUAD (blue solid lines) with a total time of $t_f = 38$ ns, a linear pulse (green dashed lines) with a total time of $t_f = 69$ ns, a $\pi$-pulse (orange dotted-dashed lines) with $t_f = 37$ ns of duration, a LA (red dots) with $t_f = 39.6$ ns. The other parameters are the same as those in Fig.~8 of the main text.}
			\label{fig:SWAP_stochastic}
		\end{figure}
		
	\end{widetext}

		\bibliography{references}
	
\end{document}